\documentstyle[twocolumn,aps,epsfig]{revtex}
\begin{document}

\draft

\title{
The Origin of the area law of the entropy of a quantum field in a black hole 
}

\author{Hyeong-Chan Kim\cite{e:kim}}
\address{           
Department of Physics, Sung Kyun Kwan University , Suwon, 440-746, Korea}
\author{Min-Ho Lee\cite{e:lee}}
\address{ Department of Physics, Kumoh University of Technology, Kumi, 
730-701, Korea}
\author{Jeong-Young Ji\cite{e:ji}}
\address{
Department of Physics Education, Seoul National University, Seoul,
151-742,  Korea}

\maketitle

\begin{abstract}
It is shown that that the area law for the entropy of a quantum field 
in the Schwarzschild black hole is due to the quantum statistics.
The entropies for one particle, a Boltzmann gas, a quantum mechanical gas
obeying Bose-Einstein or Fermi-Dirac statistics, and a quantum field
in the Schwarzschild black hole are calculated 
using the microcanonical ensemble approach and the brick wall method.
The area law holds only when the effect of quantum statistics is dominated. 
\end{abstract}


\narrowtext

%%%%%%%%%%%%%%%%%%%%%%%%%%%%%%%%%
%\section{Introduction}
%%%%%%%%%%%%%%%%%%%%%%%%%%%%%%%%%

Since the discovery of the black hole entropy 
$S^{\rm BH} = A_{\rm H}/(4 l_P^2)$ by
Bekenstein~\cite{bekenstein} and Hawking~\cite{hawking},
Euclidean path-integral approach~\cite{gibbons,york,brown,braden}, 
microcanonical functional integral approach~\cite{brown2,martinez} 
were used to derive black hole thermodynamical properties.

The statistical-mechanical foundation of the entropy of a black hole 
was discussed in connection with the information 
approach~\cite{bekenstein,zurek,thorne} and with the entanglement 
entropy \cite{thooft,bom,frolov93,sre}.
The entanglement entropy is proportional to the area of the horizon
but diverges due to the presence of arbitrary high modes near the horizon. 
To resolve this problem, 
many renormalization or regularization 
methods~\cite{susskind,fursave,spkim,frolov93,barvinsky,mhlee} 
were introduced and these methods were discussed and compared 
in Ref.~\cite{fro,lib}. 

It is generally believed that the entropy of a quantum field in a black hole 
is proportional to the area of the black hole horizon. The presence of 
the event horizon makes the one-particle number of states divergent
and singles out only the area dependence of the number of states. 
But it does not determine the functional form of the entropy on area
and if we use the result of Padmanabhan~\cite{pad} the entropy
of one-particle system is not proportional to the area 
but dependent on the {\rm logarithm} of the area.
Therefore the following question arises: 
What is the origin of the area law for the entropy of the quantum field?
There are many different properties between the classical one-particle system 
and quantum field system. The purpose of the present Letter is
to examine which aspect of quantum fields gives the area law.

For this purpose,
we adopt the micro-canonical approach and the brick-wall method.
First we reproduce the Padmanabhan's result to calculate the entropy of
the classical one-particle system. This result can be applied to the quantum 
one-particle system. Secondly, we consider the many particle systems
with the energy and the particle number fixed. Here we examine
the differences of statistical behavior 
among the Boltzmann statistics, Bose-Einstein statistics
and Fermi-Dirac statistics. 
Finally we remove the constraint on the particle number and
we take ensemble sum over all particle number states
for the quantum-field statistics. 

%%%%%%%%%%%%%%%%%%%%%%%%%%%%%%%%%
%\section{Classical entropy of a Schwarzschild blackhole}
%%%%%%%%%%%%%%%%%%%%%%%%%%%%%%%%%

The line element of Schwarzschild black hole is described by
\begin{eqnarray}
ds^2 &=& g_{tt}(r) dt^2
	 + g_{rr}(r) dr^2 + r^2 d^2\Omega_2
\label{metric} ,
\end{eqnarray}
where $d\Omega_2$ is the metric of the unit 2-sphere, and
$g_{tt} = -1/g_{rr} = -(1-2M/r)$.
Since $\xi^\alpha =(1,0,0,0)$ is a timelike Killing vector,
we can define a covariant, conserved energy  of a point 
particle to be $ H(p,x) = \xi^\alpha p_\alpha$.

Before turning on to a system of many particles, we first
consider a particle with the mass $\mu$ in a static spherical box
around a Schwarzschild black hole.
Here the inner and the outer  radii of the box 
are $2M +h$ and $L$, respectively.
The phase volume of the classical particle for a fixed energy $E$
is the volume of a hypersurface satisfying $ H(p,x) = E$, or 
\begin{eqnarray}   \label{constraint1}
\frac{p_r^2}{g_{rr}} + \frac{p_\theta^2}{g_{\theta \theta}} +
	\frac{p_\phi^2}{g_{\phi \phi}} = \frac{E^2}{-g_{tt}} - \mu^2 .
\end{eqnarray}

On the other hand, for a quantum particle which satisfies
\begin{eqnarray} \label{Psi:0}
\left[ \nabla_\mu \nabla^\mu - \mu^2 \right] \Psi =0   ,
\end{eqnarray}
using the WKB solution
\begin{eqnarray} \label{Psi}
\Psi = e^{-i  E t + i m \phi + i S(r,\theta) },
\end{eqnarray}
we have the same constraint equation (\ref{constraint1}) with
$p_r = \partial S/\partial r, p_\theta= \partial S/\partial \theta$.
In a box normalization with an appropriate boundary condition
like as in the brick wall method~\cite{thooft}, a discrete momentum 
eigenvalue corresponds to one quantum state per unit volume. 
Then the sum over the quantum states can be rewritten as the 
integral over the space and the momentum.

Thus both in the classical and the quantum systems,
the volume of the hypersurface (\ref{constraint1}) 
$ g_1(E) = \int d^3 p d^3 x \delta(E- H(p,x)) $
gives the number of states for a given energy and
can be obtained by $\partial \Gamma/\partial E$, where
$\Gamma(E) =  \int d^3 p d^3 x \theta(E- H(p,x)) $:
\begin{eqnarray}
\Gamma(E) =  \frac{4 \pi}{3} \int_{box} d^3x \sqrt{\gamma}
		\left( \frac{E^2}{-g_{tt}} - \mu^2 \right)^{3/2} ,
\end{eqnarray}
where $\gamma$ is the determinant of the spatial part of the metric.
When the inner wall of the box approach to the horizon 
($h \rightarrow 0 $),
in the leading order we have
\begin{eqnarray}
\Gamma(E) \sim \frac{2 \pi}{3} \frac{A}{\kappa^3 \epsilon^2}E^3 ,
		\\
g_1(E) \sim 2 \pi \frac{A}{\kappa^3 \epsilon^2}E^2  , \label{g:A}
\end{eqnarray}
where $A$ is the area of the inner wall of the box,
$\kappa=  1/(4M)$ is the surface gravity, at the horizon and 
$\epsilon$ is the proper distance from the horizon to the box
\begin{eqnarray} \label{epsilon}
\epsilon = \int_{2M}^{2M+ h} dr \sqrt{g_{rr}} = 2 \sqrt{2 M h} .
\end{eqnarray}
In fact, the one-particle phase volume
of the system can be consulted in Ref. \cite{pad}. 
Here we rederived it to introduce our notations.

>From the expression (\ref{g:A}) and 
the entropy formula $S(E) = \ln g_1(E) $
 we get the leading behavior 
of the entropy of the particle
\begin{eqnarray}
S \simeq \ln \left( \frac{2\pi}{\kappa^3} 
\frac{A}{\epsilon^2}E^2\right) .
\label{S:1}
\end{eqnarray}
As seen from this equation the entropy of the one-particle system 
is proportional to the logarithm of the area of the black hole area. 

Now let us  extend our result to many particle systems. 
The number of accessible states with total energy $\cal E$
and total number $N$ is~\cite{huang}
\begin{eqnarray}
g_N({\cal E}) = \sum_{\{n_i\}} W\{n_i\} ,
\end{eqnarray}
where the sum extends over all sets of integers $\{ n_i \}$ satisfying
the conditions
\begin{eqnarray} \label{cond}
{\cal E}= \sum_i E_i n_i , ~ N= \sum_i n_i .
\end{eqnarray}
Here $W\{n_i\} $ is the number of states of the system corresponding to 
the set of occupation numbers $\{n_i\}$, is given by
\begin{eqnarray}
W\{n_i\}  =
\left\{ \begin{array}{l}
\displaystyle
\prod_i \frac{g_1^{n_i}(E_i) }{n_i !}  
~\mbox{(Boltzmann)}, \\
\displaystyle
\prod_{i}{\frac{(n_i
			+g_1(E_i)-1)! }{n_i!(g_1(E_i)-1)! }}  
~\mbox{(Bose)},  \label{W:bose} \\
\displaystyle
\prod_{i}{\frac{g_1(E_i)! }{n_i!(g_1(E_i)-n_i)! }}
~\mbox{(Fermi)} \label{W:fermi}
\end{array} \right.
\end{eqnarray}
for Boltzmann statistics, Bose-Einstein statistics, 
and Fermi-Dirac statistics, respectively.

In fact, $g_N({\cal E})$  is quite well approximated by $W\{\bar{n}_i \} $ 
where $\{\bar{n}_i \}$ is a set of occupation numbers that maximize 
$W\{n_i\}$ subject to (\ref{cond})
\begin{eqnarray} \label{n:zbeta}
\bar{n}_i =
\left\{ \begin{array}{l}
		 g_1(E_i) z e^{-\beta  E_i} 
~\mbox{(Boltzmann)} ,\\
\displaystyle
  \frac{g_1(E_i)}{z^{-1} e^{\beta E_i} \mp 1 } 
~\mbox{ (Bose and Fermi)} .
		\end{array} \right.
\label{nbar}
\end{eqnarray}
Here the Lagrangian multipliers $z$ and $\beta$
are determined by the two conditions of Eq.~(\ref{cond}).
Then the entropy $S = \ln W\{ \bar{n}_i \}$ is given 
by using the Stirling's formula:
\begin{eqnarray}
S = 
\left\{ \begin{array}{l}
\displaystyle
\sum_i \bar{n}_i \log [g_1(E_i) \sqrt{n_i} ] 
~\mbox{(Boltzmann)}, \\
\displaystyle
\sum_i 
\left[ \bar{n}_i \ln \left(1+ 
\frac{ g_1(E_i)}{\bar{n}_i } \right) 
+ g_1(E_i) \ln\left(1+ \frac{\bar{n}_i }{g_1(E_i)}
\right) \right] 
~\mbox{(Bose)}  , \\
\displaystyle
 \sum_i 
\left[ \bar{n}_i \ln \left(\frac{
g_1(E_i)-1}{\bar{n}_i }\right) 
- g_1(E_i) \ln\left(1- \frac{\bar{n}_i }{g_1(E_i)}
\right) \right] ~\mbox{(Fermi)} .
\end{array} \right.
\end{eqnarray}
The explicit calculation using (\ref{g:A}) and (\ref{n:zbeta}) gives
the entropy
\begin{eqnarray} 
\label{entropy}
S  = 
\left\{ \begin{array}{l}
\displaystyle
\frac{N}{2} \left[ 3 + 2 \ln \left( \frac{2 \alpha {\cal E}^3}{27 N^4} 
		\right) \right]  
~\mbox{(Boltzmann)}, \\
\displaystyle
\frac{2 \alpha}{\beta^3} 
	\left[4 f_4(z) -  f_3(z) \ln z \right] 
~\mbox{(Bose and Fermi)} .
\end{array} \right.
\label{e16}
  \end{eqnarray}
where  $ \alpha = \frac{2\pi A}{\kappa^3 \epsilon^2}$
and
$f_n(z) = \frac{1}{(n-1)! }  \int_0^\infty dx x^{n-1}  / (e^x/z \mp 1) $ 
(Bose and Fermi)  
can be written as
\begin{eqnarray}
 f_n(z) =
\left\{ \begin{array}{l}
\displaystyle
 \sum_{l=1}^\infty \frac{z^l}{l^n}
\mbox{ for } 0 \leq z \leq 1    \mbox{ (Bose)}, \\
\displaystyle
-\sum_{l=1}^\infty \frac{(-z)^l}{l^n}  
\mbox{ for }  0 \leq z \leq 1
\mbox{ (Fermi)}\\
\displaystyle
\frac{1}{n!}  \sum_{i=0}^{[n/2]} {_n}C_{2i} (\ln z)^{n-2i} I_{2i}
		+ O\left(\frac{1}{z}\right)  \\
\hspace{2.3cm} \mbox{ for }  z \gg 1 ,
\mbox{ (Fermi)} 
 \end{array} \right.
\end{eqnarray}
using the Sommerfeld expansion for large $z$.
Here $[x]$ denotes the greatest integer which is not greater than $x$,
and $I_{n} = (n-1)!(2n)(1-2^{1-n})\zeta(n)$.

Let us discuss the result of Bose statistics, 
to see the change of the entropy behavior 
from the logarithm dependent one to the linear one 
of the area of the black hole. 
Note that  Eq.~(\ref{constraint1}) gives restriction that the 
ground state energy must be greater than $E_0 = \mu \epsilon/(4M)$.
In fact this is extremely small quantity so
we can set it effectively to zero. 
With this choice the  constraint equations (\ref{cond})
give the following equation for $\beta, z$.
\begin{eqnarray}
\int dE  \frac{g_1(E)}{e^{\beta E}/z -1} + n_0 = N , ~
\int dE \frac{E g_1(E)}{e^{\beta E}/z -1}  = {\cal E}  ,
\end{eqnarray}
where we isolated 
the average number of particles in the zero energy state
$n_0= z/(1-z)$ which is the zero energy divergence 
in Eq. (\ref{nbar}) at $z=1$.  
These  constraint equations become
\begin{eqnarray}  \label{z:EN}
\frac{f_3(z)}{\beta^3}+ \frac{n_0}{2 \alpha} = \frac{N}{2 \alpha}, ~ 
\frac{f_4(z)}{\beta^4} = \frac{{\cal E}}{6 \alpha} ,
\label{e19}
\end{eqnarray}

For a small $z$, we can approximate $f_n(z) = z+ z^2/2^n $
and we have
\begin{eqnarray}
z & \simeq &  \frac{27 N^4}{2 \alpha {\cal E}^3 },  \\
\beta & \simeq & \left(1-\frac{27 N^4}{32 \alpha {\cal E}^3} \right) 
		\frac{3N}{{\cal E}} .
\label{b:NE}
\end{eqnarray}
These equations hold only for 
a relatively small number of particles 
compared to the phase volume of one-particle system with energy
${\cal E}/N$ 
(Note that  the $z \ll 1$ means 
$N/[\alpha ({\cal E}/N)^3] \ll 1$).  
Then the entropy is given by 
\begin{eqnarray}
S \simeq  N \ln \frac{4 \pi A {\cal E}^3}{27
		\kappa^3 \epsilon^2 N^4} ,
\label{S:smallz}
\end{eqnarray}
which is quite similar to the Boltzmann gas in (\ref{e16}).
If we consider the on-shell entropy by using
Eq.~(\ref{b:NE}) and $\beta \kappa = 2 \pi$, we get
\begin{eqnarray}
S \simeq N \ln \frac{ A \left(1- 27N^4/32 \alpha {\cal E}^3 \right)
		}{2 \pi^2 N \epsilon^2} ,
\end{eqnarray}
which is proportional to the logarithm of the area.

\begin{figure}[htb]
\centerline{\epsfig{figure=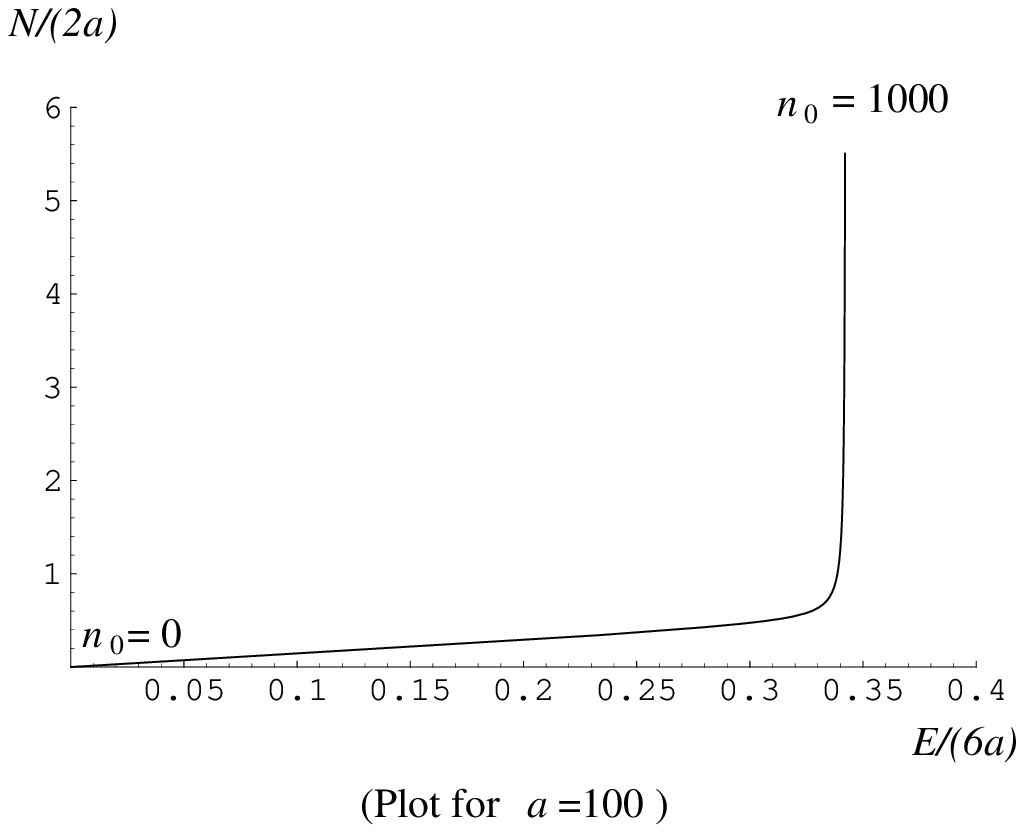,width=100mm,angle=0}}
\caption{
${\cal E}$-$N$ diagram. 
The line is depicted for $\beta=0.75$ with $\alpha =100$. 
The number
of particles in the ground states varies from $0$ to $1000$.
}
\label{fig2}
\end{figure}

For $z \sim 1 ( n_0 \gg 1 )$, from (\ref{z:EN}) we have 
\begin{eqnarray} 
\label{E:max}
{\cal E} = \frac{6 \alpha \zeta(4)}{\beta^4} , ~
N = n_0 + 2 \alpha \frac{\zeta(3)}{\beta^3} ,
\end{eqnarray} 
where we used the approximation  $f_i(z \sim 1) \simeq \zeta(i)$. 
Here the energy value in (\ref{E:max}) is the maximum energy the system 
can have for a fixed $\beta$.  
On the other hand the number of particles can grow indefinitely
due to the presence of the zero energy particles as can be seen in Fig.~1.
This phenomenon that the zero energy particles pile up in the ground
state is the Bose-Einstein condensation.
Now the entropy is given by
\begin{eqnarray}
S = \frac{16 \pi \zeta(4)}{\beta^3 \kappa^3 \epsilon^2} A .
\label{S:z1}
\end{eqnarray}
Here we note that the black hole entropy in this limit obeys the area law.

Now let us consider the Fermi statistics.
The constraint equations (\ref{cond}) can be rewritten as
\begin{eqnarray} \label{fer:constraint}
\int dE  \frac{g(E)}{e^{\beta E}/z +1}  = N, ~
\int dE \frac{E g(E)}{e^{\beta E}/z +1}  = {\cal E},
\end{eqnarray}
and after some algebra we get
\begin{eqnarray} \label{con1}
\frac{f_3(z)}{\beta^3}  = \frac{N}{2 \alpha},
\frac{f_4(z)}{\beta^4} = \frac{{\cal E}}{6 \alpha}.
\label{e28}
\end{eqnarray}
For small $z$  the solution of  these equations are exactly the same
as those of Bose statistics (\ref{S:smallz}). 
On the other hand,  for  large $z$ we get
\begin{eqnarray}\label{fer:EN}
{\cal E} &=& \frac{\alpha}{4} \left[ \left( \frac{\ln z}{\beta}\right)^4  
+ 6 \frac{I_2}{\beta^2}  \left( \frac{\ln z}{\beta}\right)^2
 +  \frac{I_4}{\beta^4} \right],
\nonumber \\
N  &=& \frac{\alpha}{3} \left[ \left( \frac{\ln z}{\beta}\right)^3 
  + 3 \frac{I_2}{\beta^2}  \frac{\ln z}{\beta}  \right].
\end{eqnarray}
Then the entropy obeys the area law again:
\begin{eqnarray}
S = \frac{\alpha}{\beta} \left(I_2  \left( \frac{ \ln z}{\beta} \right)^2 
+ \frac{I_4}{3}
\label{S:FD}
\right).
\end{eqnarray}

All the previous results are obtained under the assumption
that the total number of particles, $N$, is constant. However,
these can be regarded as quantum mechanical limits 
because the Fock space of quantum field theory is spanned 
by a complete set of particle number states. 
Therefore when we count the accessible states
in the {\it microcanonical ensemble} for a quantum field
we should consider all possible number states,
with the energy being fixed.
Then it is natural to vary the particle numbers and
we fix the energy only: 
\begin{eqnarray} \label{con}
{\cal E} = \sum_i n_i E_i &=& 6 \alpha \frac{f_4(z)}{\beta^4} 
\end{eqnarray}
both for Bose and Fermi statistics.
The number of states $g({\cal E})$ for this ensemble can be obtained 
by summing $g_N({\cal E})$  over all possible $N$:
\begin{eqnarray} \label{g(E)}
g({\cal E}) = \sum_{N=0}^{\infty} g_N({\cal E}).
\end{eqnarray}
Now we have a good approximate $g({\cal E}) \simeq g_{\bar{N}}({\cal E})$, 
where we can find $\bar{N}$ applying the maximal entropy principle to $g_N (E)$.
By varying (\ref{con}) with respect to $z$ and $\beta$ with ${\cal E}$ fixed,
we have
\begin{eqnarray}
\delta \beta = \frac{\beta}{4 z} \frac{f_3(z)}{f_4(z)} \delta z,
\end{eqnarray}
where we used $f'_n(z) = f_{n-1}(z)/z$.
Further, from the variation of $g_N({\cal E})$ we have 
\begin{eqnarray}
\delta \ln g_N({\cal E}) = \frac{\alpha}{\beta^3 z} \delta z 
	F(z) \ln z= 0 ,
\label{e35}
\end{eqnarray} 
where $F(z) =  3 f_3^2 (z)/4 f_4(z) - f_2(z) $. 
Since $F(z)$ is always negative for $z>0$ and 
$F(z \rightarrow  0) \sim z $,  
$z=1$ is the only solution of (\ref{e35}).
Then the distribution of particles which satisfies (\ref{con})
becomes
\begin{eqnarray} \label{n}
\bar{n}_i = \frac{g_1(E_i)}{ e^{\beta E_i} \mp 1 }
\mbox{ (Bose and Fermi) }.
\label{n:max}
\end{eqnarray}
This result  can be obtained by applying the maximal entropy principle 
directly to  Eq.~(\ref{W:bose})  
with the constraint~(\ref{con}).
The entropy now is given by
\begin{eqnarray}
S_{\cal E} = \frac{16 \pi f_4(1)}{\beta^3 \kappa^3 \epsilon^2} A,
\label{S:QF}
\end{eqnarray}
where $f_4(1) = \zeta(4)$ for Bose statistics and $f_4(1) = (7/8) \cdot \zeta(4)$
for the Fermi statistics and 
the suffix ${\cal E}$ was introduced to stress
that we have fixed only the energy by the constraint (\ref{con}).
Further it can be verified that  $\partial S_{\cal E} / \partial {\cal E} = \beta$
using (\ref{e19}) and (\ref{e28}). 
If we consider the equilibrium with the black hole of the inverse temperature 
$\beta = \beta_{\rm BH} = 2 \pi /\kappa$,
the entropy is given by 
\begin{eqnarray}
S_{\cal E} = \frac{2 f_4(1)}{\pi^2 \epsilon^2} A .
\end{eqnarray}
The relative ratio between the entropy of Fermi gas and the 
entropy of Bose gas in equilibrium with the black hole is $7/8$.

%%%%%%%%%%%%%%%%%%%%%%%%%%%%
%\section{Summary and Discussions}
%%%%%%%%%%%%%%%%%%%%%%%%%%%%

Now we can answer to our main question about area law of the entropy.
As can be seen in the cases of large particle number limit 
of quantum mechanical systems [see (\ref{S:z1}) (Bose-Einstein)
and (\ref{S:FD}) (Fermi-Dirac)] or quantum field case [see (\ref{S:QF})],
the main contribution of the entropy is proportional to the area.
On the other hand, if the number of particles is small or if the system
obeys Boltzmann statistics [see (\ref{S:1}) for one-particle, (\ref{e16}) for 
Boltzmann gas, and (\ref{S:smallz}) for small number limit of quantum particles],
the entropy is dependent on the logarithm of the area.  Therefore we can
conclude that the area law for the entropy of a quantum field 
in the Schwarzschild black hole is due to the quantum statistics.
The role of large particle number is to expose the effect of quantum statistics.
In the case of quantum fields, the maximal entropy principle automatically 
selects the states whose particle numbers give the area law [see (\ref{n:max})].

We would like to thank Prof. Yoonbai Kim and Prof. Philliar Oh
for stimulating discussions and Prof. R. Ruffini for his encouragement.
H.C.K. and J.Y.J. are supported by the Korean Science and Engineering 
Foundation  for the post-doctoral fellowships 
and the Center for Theoretical Physics (SNU).
M.H.L. is supported by the Center for Molecular Sciences (KAIST).


\begin{thebibliography}{8}
\bibitem[*]{e:kim} Electronic address:hckim@chep5.kaist.ac.kr
\bibitem[\dag]{e:lee} Electronic address:mhlee@chep5.kaist.ac.kr
\bibitem[\ddag]{e:ji} Electronic address:jyji@phyb.snu.ac.kr

{byline}
\bibitem{bekenstein} J. D. Bekenstein, Phys. Rev. D {\bf 7}, 2333, (1973);
        J. M. Bardeen, B. Carter, and Hawking, Commun. math. Phys. {\bf 31},
        161, (1973).
\bibitem{hawking}
        S. W. Hawking, Commun. Math. Phys. {\bf 43}, 219 (1975).
\bibitem{gibbons}
        G. W. Gibbons and S. W. Hawking, Phys. Rev. D {\bf 15}, 2752 (1976).
\bibitem{york} J. W. York Phys. Rev. D {\bf 33}, 2029 (1986).
\bibitem{brown} J. D. Brown, G. L. Comer, E. A. Martinetz, J. Malmed,
        B. F. Whiting and J. W. York, Class. Quant. Grav. {\bf 7}, 1433 (1990).
\bibitem{braden} H. W. Braden, J. D. Brown, B. F. Whiting, and J. W. York,
        Phys. Rev. D {\bf 42}, 3376 (1990).
\bibitem{brown2} J. D. Brown and J. W. York, Jr.,  Phys. Rev. D {\bf 47},
                1407 (1993).
\bibitem{martinez} E. A. Martinez, Phys. Rev. D {\bf 51}, 5732 (1995).
\bibitem{zurek} W. H. Zurek and K. S. Thorne, Phys. Rev. Lett. {\bf 54}, 2171
                (1985).
\bibitem{thorne} K. S. Thorne, R. H. Price, and D. A. Macdonald. 
        ``Black Holes: The Membrane Paradigm" (Yale Univ Press, NewHaven 
        and London) (1986).
\bibitem{thooft} G. 't Hooft, Nucl. Phys. B {\bf 256},  727 (1985).
\bibitem{bom} L. Bombelli, R. K. Koul, J. Lee, and R. D. Sorkin, Phys. Rev. D
        {\bf 34}, 373 (1986).
\bibitem{frolov93} V. P. Frolov and I. Novikov, Phys. Rev. D {\bf 51}, 618 (1993).
\bibitem{sre} M. Srednicki, Phys. Rev. Lett. {\bf 71}, 666 (1993).
\bibitem{susskind} L. Susskind and J. Uglum,  Phys. Rev. D {\bf 50},
                2700 (1995).
\bibitem{fursave} D. V. Fursave and Solodukhin, "On one-loop renormalization
        of black hole entropy" Preprint hep-th/9412020.
\bibitem{spkim} S. P. Kim, S. K. Kim, K.-S. Soh, and J. H. Yee, 
	Phys. Rev. D {\bf 55} 2159 (1997).
\bibitem{barvinsky} A. O. Barvinsky, V. P. Frolov, A. I. Zelnikov, Phys. Rev. D
        {\bf 51}, 1741 (1995). 
\bibitem{mhlee} M.-H. Lee and J. K. Kim, Phys. Lett. A {\bf 212}, 323 (1996).
\bibitem{fro} V. P. Frolov, D. V. Fursaev, and A. I. Zelnikov, Phys. Rev. D
        {\bf 54} 2711 (1996).
\bibitem{lib} S. Liberati, "Problems in Black Hole Entropy interpretation",
        Talk presented at the "Fourth Italian-Korean meeting on Relativistic 
        Astrophysics", Rome-Gran Sasso - Pescara, July 9-15, (1995).
\bibitem{pad} T. Padamanabhan, Phys. Lett. A {\bf 136}, 203 (1989).
\bibitem{huang} K. Huang, "Statistical Mechanics", 2nd Eds. 
		(John Wiley \& Sons) (1986).
\end{thebibliography}
\end{document}